\documentclass[twocolumn]{aastex631}
\usepackage{multirow}
\usepackage{wasysym}
\usepackage{enumerate}
\usepackage{amssymb} 

\shorttitle{MIRI Persistence}
\shortauthors{Vasan et al.}

\begin{document}
\title{A Characterization of JWST MIRI Detector Persistence \& Implications for High-Contrast Imaging}


\correspondingauthor{Alisha Vasan}
\email{alishakv@umich.edu}

\author[0009-0007-2342-3996]{Alisha Vasan}
\affiliation{Department of Astronomy, University of Michigan, Ann Arbor, MI 48109, USA}

\author[0000-0002-9521-9798]{Mary Anne Limbach}
\affiliation{Department of Astronomy, University of Michigan, Ann Arbor, MI 48109, USA}

\author[0000-0001-7246-5438]{Andrew Vanderburg}
\affiliation{Department of Physics and Kavli Institute for Astrophysics and Space Research, Massachusetts Institute of Technology, Cambridge, MA 02139, USA}

\author[0000-0001-5831-9530]{Rachel Bowens-Rubin}
\affiliation{Department of Astronomy, University of Michigan, Ann Arbor, MI 48109, USA}
\affiliation{Eureka Scientific Inc., 2542 Delmar Ave., Suite 100, Oakland, CA 94602, USA}

\author[0000-0002-7352-7941]{Kevin B. Stevenson}
\affiliation{JHU Applied Physics Laboratory, 11100 Johns Hopkins Rd, Laurel, MD 20723, USA}

\begin{abstract}
The JWST MIRI detector exhibits { a flux deficit} persistence, but its timescales and impacts remain largely uncharacterized, particularly at the longest imaging wavelengths.  In this study, we analyze full-field MIRI imager observations at 21\,$\mu$m (F2100W) to quantify detector persistence following a saturation event by a bright (K = 5.65 mag) nearby (8.12\,$\pm$\,0.04 pc) mid M-dwarf star, IRAS 21500+5903. { Unlike typical persistence that appears as excess flux, this effect presents as a flux deficit in pixels previously illuminated by the saturating or near saturating source.} We measure persistence at two post-saturation epochs: shortly after saturation (11.6 minutes) and an hour later (1.39 hours). Immediately after the saturation event, we detect a persistence level of $1.69 \pm 0.10$\%. By fitting a Bayesian exponential decay model to the two epochs, we estimate that persistence decreases to one-tenth of its initial value after \(5.16^{+1.49}_{-0.94}\) hours. We examine the implications of persistence for MIRI high-contrast imaging using the imager (not coronagraphy). Specifically, we discuss how MIRI detector persistence can produce false-positive exoplanet signals in direct imaging surveys, as well as degrade PSF subtraction, particularly at small inner working angles. We also outline mitigation strategies to avoid these impacts in future observations.

\end{abstract}
\section{Introduction}

As JWST enters full science operations, understanding the behavior and performance of its instruments is essential for optimizing data quality and scientific return. 
The Mid-Infrared Instrument (MIRI) imager \citep{2023PASP..135d8003W} on JWST is enabling groundbreaking observations of galaxies \citep{2023arXiv230602465E,2023ApJ...953L..29L}, debris disks \citep{2023NatAs...7..790G}, exoplanets \citep{2024ApJ...962L..32M,2025ApJ...984L..28L}, and star forming regions \citep{2024A&A...685A..73H} with its unprecedented mid-infrared sensitivity. With these powerful capabilities, however, come subtle instrumental effects that must be carefully characterized, especially as MIRI is used to make increasingly precise and sensitive measurements. 

One of the challenges is that the MIRI detector exhibits a known issue called \textit{persistence}\footnote{\scriptsize \url{https://jwst-docs.stsci.edu/jwst-mid-infrared-instrument/miri-instrumentation/miri-detector-overview/miri-detector-performance}}. Persistence is an artifact common in infrared detectors { that} can introduce false structures into images, bias photometric measurements, { contaminate faint source photometry, and degrade PSF subtraction performance}, particularly in sequences where both bright and faint targets are observed together or in close succession.

{It is important to distinguish between different types of persistence effects. Most detector persistence presents as excess flux where charge from a bright source is retained in the detector and released slowly over time, creating residual signals in subsequent exposures \citep{2019JATIS...5c6004T}. However, MIRI exhibits a less common form of persistence, where very bright sources create a negative offset in the pixels they have illuminated, resulting in a flux deficit rather than the more typical persistence excess \citep{2015PASP..127..584R,2015PASP..127..665R}. The physical mechanism underlying this flux-deficit persistence is not yet fully understood, but it appears to be linked specifically to near or hard detector saturation events.}

While previous studies have found persistence effects to be minimal in some of MIRI’s shorter-wavelength imaging bands \citep[$\lambda < 12.8\,\mu$m;][]{2023PASP..135g5004M,2024A&A...689A...5D}, the impact on faint sources in the context of high-contrast imaging, especially where both faint and bright sources coexist in the image, has not been well characterized. Furthermore, little is known about persistence behavior in MIRI’s longest ($>18\mu$m) imaging wavelengths, including the F1800W, F2100W and F2550W filters, { where flux-deficit persistence signals appear more prominent and likely to contaminate measurements}.

In this paper, we measure the persistence decay timescale in the MIRI F2100W filter by analyzing imaging observations of the bright star IRAS~21500+5903, taken minutes to hours after a saturation event. We use two datasets obtained at different time intervals following saturation to quantify the decay of the residual persistence signal. These results provide empirical constraints on how long persistence remains detectable, offering practical guidance for planning MIRI observations and mitigating persistence-related artifacts in future science programs. The layout of this paper is as follows. Section~\ref{sec:obs} describes the data used. Section~\ref{sec:Analysis} outlines our analysis approach and presents the measured MIRI F2100W persistence decay rate. In Section~\ref{discussion}, we discuss how persistence on the MIRI imager impacts high-contrast imaging and explore mitigation strategies. Finally, Section~\ref{summary} summarizes our findings.

\section{Observations} \label{sec:obs}

\begin{figure*}
    \centering   \includegraphics[width=1\linewidth]{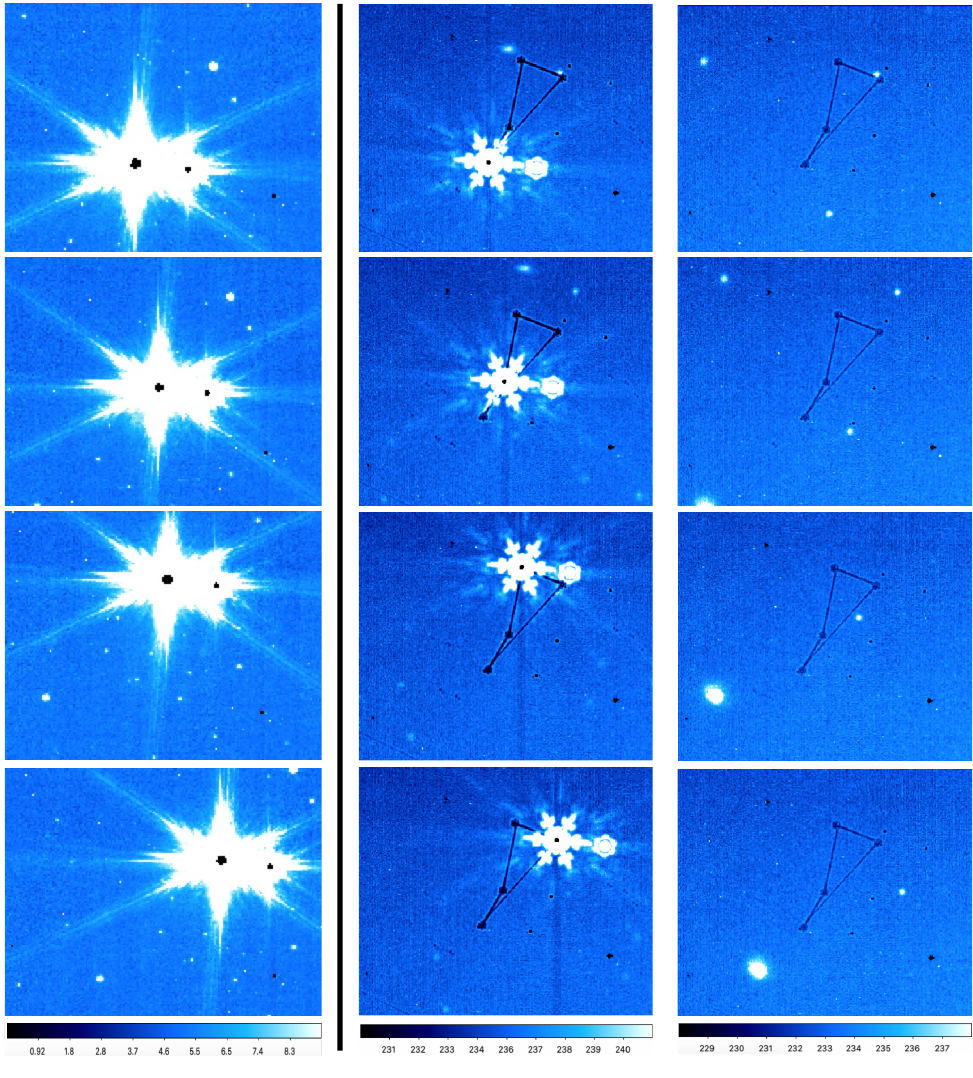}
    \caption{{ Saturation event occurs when imaging WD~2151+591 in the F770W band (left).} Sequential dither position images with MIRI at F2100W of {WD~2151+591} ({ middle}) and WD~1748+708 (right) illustrate the {flux-deficit} persistence caused by saturation from the bright M-dwarf IRAS~21500+5903 in the WD~2151+591 field. This source saturated the detector most severely in 7.7\,$\mu$m (F770W; { left}) imaging taken 12\, minutes prior to the imaging sequence on the { middle} and 1.39\, hours prior to the sequence on the right. {The flux-deficit} residual trace from the M-dwarf is visible between dither positions as it moves across the MIRI detector. Dither pattern used for this observation was {\it cycling} with dither points $1-4$ (top to bottom). 
}
    \label{fig:enter-label1}
\end{figure*}

\begin{deluxetable}{ccccc}

\tablecaption{Relevant information concerning observations. \label{tab:observation information}}
\tablehead{
\colhead{Target} & \colhead{Obs no.} & \colhead{Filter} & \colhead{Dithers} & \colhead{Total time [s]}
}
\startdata
WD~2151 & 9 & F770W & 4 & 55.5\\
WD~2151 & 9 & F1800W & 4 & 710.412\\
WD~2151 & 9 & F2100W & 4 & 277.504\\
\hline
WD~1748 & 3 & F770W & 4 &55.5 \\
WD~1748 & 3 & F1800W & 4 &710.412 \\
WD~1748 & 3 & F2100W & 4 & 277.504\\
\enddata
\tablecomments{The specific observations analyzed can be accessed via \dataset[doi: 10.17909/cjkx-kp07]{https://doi.org/10.17909/cjkx-kp07}}.
\end{deluxetable}

Our study uses data from the JWST Cycle 2 survey program \#4403 \citep{2023jwst.prop.4403M,Limbach2024}. This survey collected 7.7, 18.0, and 21.0\,$\mu$m MIRI imaging on about 20 nearby white dwarfs. One of the white dwarfs imaged in this survey, 
{WD~2151+591}, had a bright main sequence companion, IRAS~21500+5903, which saturated the detector. 
IRAS~21500+5903 is located at a distance of 8.12 ± 0.04 parsecs based on a Gaia DR2 parallax measurement of 123.06 ± 0.59 milliarcseconds \citep{2018yCat.1345....0G}. The star is an M4V dwarf companion, has a K-band magnitude of 5.65, and a WISE band W4 (22~$\mu$m) magnitude of 5.14. 
 
The imaging sequence progressed from the shortest (7.7\,$\mu$m) to the longest (21.0\,$\mu$m) wavelength. Saturation was most severe in the initial 7.7\,$\mu$m image, leading to significant persistence in the 21.0\,$\mu$m images taken 11.6 minutes later. This persistence was still present 1.39 hours later in a white dwarf system (WD~1748+708), which was imaged with MIRI using the same sequence as part of this survey program. At the start of our analysis in mid-2024, we also searched the MAST archive for additional datasets exhibiting significant persistence in the F2100W and F2550W bands that could be used to characterize MIRI imager persistence at the longest wavelengths. However, we did not identify any suitable datasets beyond the imaging set analyzed in this study.
{ WD~2151+591} and WD~1748+708 were observed on September 20, 2023.

The 21\,$\mu$m imaging observations took place 1.39 hours apart and used the fast readout mode and a four-point cycling dither (starting at point 1). All imaging used in this study was conducted in the F2100W filter, with individual dithers consisting of 5~integrations per exposure and 12 groups per integration for a total of 177.6 sec observations time per dither. {A complete list of data files and pipeline reduction version (using jwst\_1252.pmap) employed in this analysis is provided in the MAST archive under program \#4403.} The images at each of the four dither positions on {WD~2151+591} and WD~1748+708 are shown in Figure \ref{fig:enter-label1}. { The persistence is visible both at the locations where the star was imaged on the detector and along its path between dither positions. We suspect that the persistence appears between dither positions, rather than from the star's initial entry into the field, because the F770W filter was selected after the telescope slewed to the target, while slewing, a lower-throughput filter was likely in place, preventing initial persistence formation.
}

\section{Analysis \& Results} \label{sec:Analysis}

\subsection{Methods}

Leveraging the 21\,$\mu$m data from these two systems, we aim to quantify  { JWST MIRI flux-deficit persistence} and measure its decay rate. While observations were conducted at multiple wavelengths (7, 18, and 21\,$\mu$m), we focus on persistence effects exclusively in the F2100W filter. This is because F2100W is the only band where precise persistence measurements are possible due to the high ``sky" background flux. The high baseline flux enables the detection of small flux deficits caused by persistence against a uniform background. The residual detector behavior that is clearly visible in F2100W images (see Figure \ref{fig:enter-label1}) is not discernible at the shortest wavelengths. { Note that persistence itself is not wavelength-dependent, but how the artifacts present in subsequent data can have wavelength dependence. For example, elevated background levels at longer wavelengths (e.g., 21\,$\mu m$ compared to  7\,$\mu m$) make small flux deficits easier to detect.}

\begin{figure}
    \centering
    \includegraphics[width=1\linewidth]{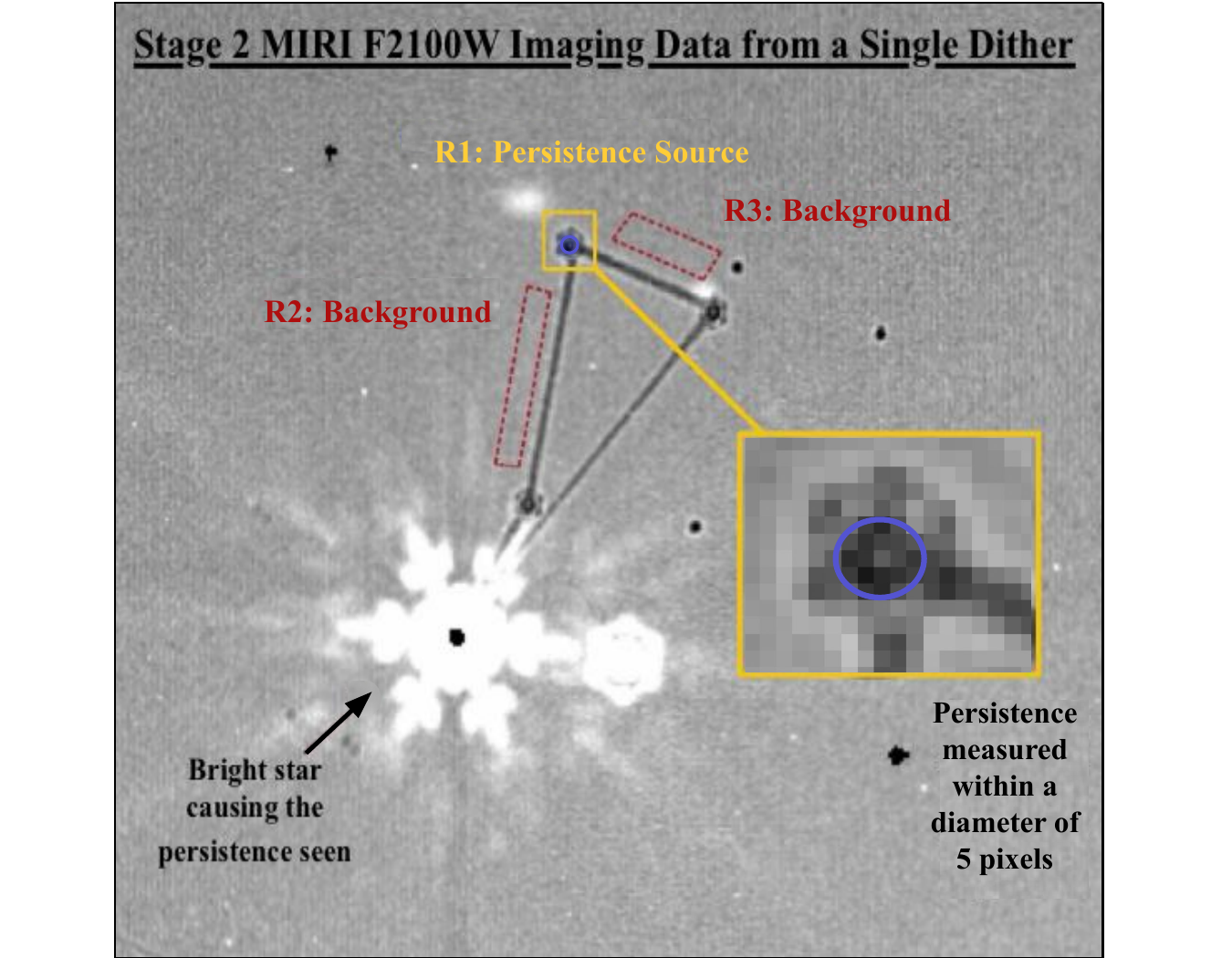}
    \caption{{\bf}Diagram illustrating the regions used for persistence measurement at one dither position: the persistence feature (R1, blue circle) and two background regions (R2 and R3, red dashed boxes). The yellow frame highlights a zoomed-in view of the persistence measurement region with a 5-pixel diameter. 
}
    \label{fig:enter-label2}
\end{figure}

Our methodology focused on measuring and comparing flux ratios between observable persistence features and their nearby background regions to quantify the magnitude of persistence. { We note that persistence can occur from both bright and saturating sources, though this particular case is linked specifically to hard detector saturation, where pixels saturate well before the end of an integration, rather than simply long integration times on a bright source.} 
We measured the flux deficit due to persistence relative to the background at each of the four dithers for both imaged systems (e.g., the eight images { on the two right columns} shown in Figure \ref{fig:enter-label1}). We computed the median flux in each region of interest at every dither position. Using the resulting persistence values from all four dithers, we then calculated the mean and standard deviation to quantify the average persistence and assess the measurement uncertainty.
Specifically, for each observation in our dataset, we performed the following measurements:
\begin{enumerate}
  \item The median flux value in the Region R1 which has a 5-pixel diameter, and corresponds to where the saturating star, IRAS~21500+5903, was centered on the detector during the initial F770W imaging. This region retains residual persistence for an extended period after the source was observed.
  \item The median flux in the nearby background regions (R2 and R3, which are illustrated in Figure \ref{fig:enter-label2}) is measured on both sides of the persistence feature. These areas provide baseline measurements of the background flux in regions unaffected by persistence.
\end{enumerate}
To quantify the magnitude of persistence, we calculated the flux deficit, defined as:
\begin{equation}
p = 1 - \frac{F_{\mathrm{persistence}}}{F_{\mathrm{background}}}
\end{equation}
where $p$ is the fractional flux deficit in the persistence region relative to the background, $F_{\mathrm{persistence}}$ is the median flux in the persistence region (R1), and $F_{\mathrm{background}}$ is the median flux in the combined background regions (R2 and R3). We measured this deficit at each dither position for both imaged systems and computed the mean and standard deviation, of the median-derived values, across the four dithers. These values were then used to calculate the decay rate of the persistence signal over time. The resulting persistence measurements are plotted in Figure \ref{fig:enter-label3}, with blue points representing the F2100W imaging taken 0.19 and 1.39 hours following the initial saturation event, with error bars indicating measurement uncertainties. In this plot, the y-axis denotes the percent deficit in background flux due to persistence as a function of elapsed time since the saturation event. In the {WD~2151+591} image taken 11.6\, minutes after the saturation event, we measure a flux deficit of $1.69 \pm 0.10\%$ due to persistence. Later, 1.39\, hours after the saturation event, the persistence has decreased to $0.99 \pm 0.10\%$. 

{ Note that our flux measurements represent time-averaged values over the 710-second integration at F2100W, rather than instantaneous persistence levels. Because the persistence signal evolves during the integration, the measured flux is effectively a combination of the true sky signal and decaying persistence and this decay in affect distorts the ramp and therefore the flux measurement. In extreme cases, persistence could lead to a curved ramp, resulting in a poor ramp fit and an inaccurate flux measurement. Here we assume this effect is negligible, as the integration time, $\mathcal{O}(10\,\mathrm{min})$, is much shorter than the total persistence decay time, $\mathcal{O}(10\,\mathrm{hr})$.}

\subsection{Decay Rate of Persistence}

\begin{figure}
    \centering    \includegraphics[width=1\linewidth]{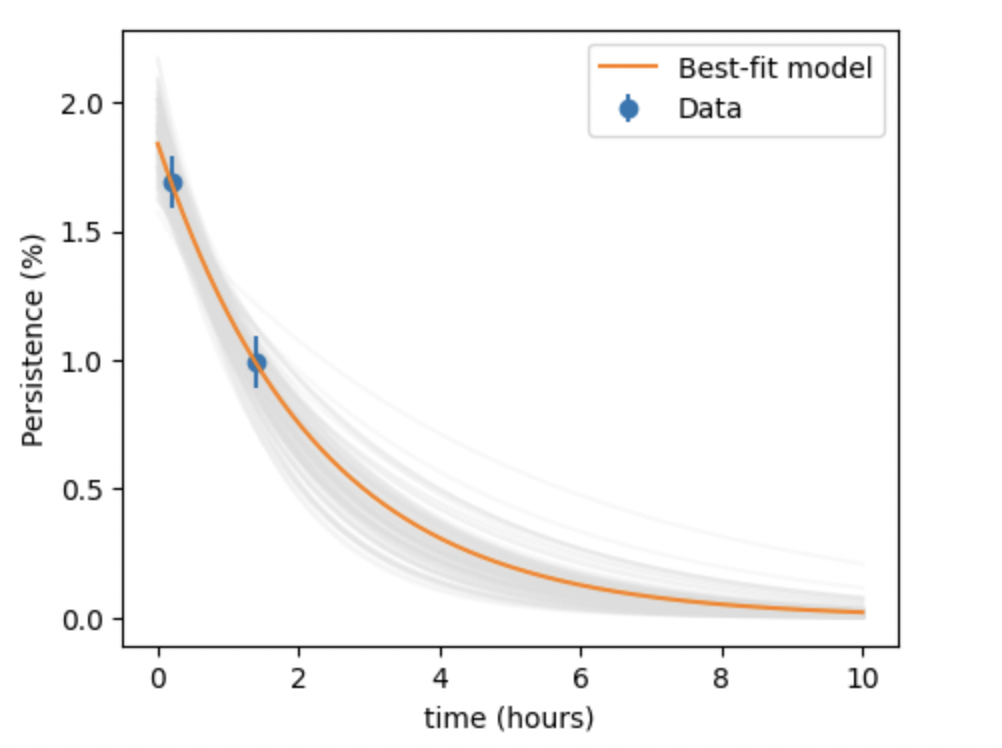}
    \caption{{\bf} Measured {flux-deficit} persistence (blue points) from the F2100W imaging taken 0.19 and 1.39 hours after the initial saturation event in the F770W band, with error bars representing the average measurement uncertainties. The orange line indicates the median exponential decay model obtained from the MCMC fit. The gray curves represent posterior draws from the MCMC sampling, illustrating the full range of decay curves consistent with the data. Based on this model, the persistence signal decreases to 1/10th of its initial value after \(5.16^{+1.49}_{-0.94}\) hours.
}
    \label{fig:enter-label3}
\end{figure}

Using our measured persistence values, we now aim to quantify the decay timescale of the persistence effect. To do this, we fit a exponential decay model to the two measured persistence values. This fit was performed using Markov Chain Monte Carlo (MCMC) sampling, which provides posterior distributions for the amplitude and decay constant, and quantifies the uncertainty in the decay timescale (Figure \ref{fig:enter-label3}).  The fit to the decay in persistence is described by the equation:  
\begin{equation}
R(t) = (1.85 \pm 0.13) \cdot e^{-0.45 \pm 0.10 \cdot t}
\end{equation} 
where \( R(t) \) represents the percentage decay rate (\%) and \( t \) is the time in hours. This exponential fit suggests that the persistence effect declines by an order of magnitude \(5.16^{+1.49}_{-0.94}\) hours after the saturation event. However, we note that \cite{2023PASP..135g5004M} suggested that there are likely multiple persistence decay timescales ranging from minutes to hours, with dependencies on the luminosity of the source and the duration of its observation. { Furthermore, characterization of the full persistence decay curve is infeasible due to data gaps caused by instrument and observatory overheads.} This suggests that the decay and manifestation of {flux-deficit} persistence are likely more complex and case-dependent than the simple exponential decay fit applied here. Nevertheless, given the limited information currently available on MIRI persistence decay rates on hour-long timescales, this work provides an initial quantification of the magnitude and duration of persistence that can be expected with MIRI.

\section{Discussion}\label{discussion}

Several programs use MIRI imaging (rather than coronagraphy) for direct imaging searches of exoplanets (and characterization of debris disks) around very nearby, bright main-sequence stars \citep{2017jwst.prop.1193B,2023NatAs...7..790G,2024jwst.prop.6122B,2024ApJ...977..277S,2025AJ....169...17B,2025jwst.prop.8581,2025AAS...24537006W}, as it has been demonstrated that this mode provides advantageous sensitivity over coronagraph in some cases \citep{2025arXiv250515995B}. However, to take advantage of this mode requires careful handling of both the brighter-fatter effect as well as persistence.





\subsection{Persistence as a Source of False Positives in High-Contrast Direct Imaging}

MIRI high-contrast imaging observations often rely on reference star differential imaging (RDI) or angular differential imaging (ADI) to remove the stellar PSF. As a result, the observational setup typically involves back-to-back imaging of stars with MIRI to maintain a stable observatory wavefront. { However, the star’s position on the detector can vary due to telescope is pointing and the accuracy of the supplied stellar position}. This misalignment can inadvertently lead to false-positive exoplanet detections in RDI-based analyses.

Consider the following scenario:  
\begin{enumerate}
\vspace{-2mm}
\item  Star 1, which is near saturation, is imaged at the exact center of the detector. 
\vspace{-2mm}
\item Shortly after, Star 2 is imaged at an offset of $sim$10 pixels from Star 1’s location. However, due to persistence, a residual negative image of Star 1 is now in the new exposure, offset by $sim$10 pixels.  
\vspace{-2mm}
\item  When RDI is performed by subtracting [Star 1] - [Star 2] to remove the stellar PSF, the persistence residual from Star 1 appears as a positive feature in the final image, mimicking the signal of a faint, close-in companion.
\end{enumerate}

\begin{figure}
    \centering    \includegraphics[width=1\linewidth]{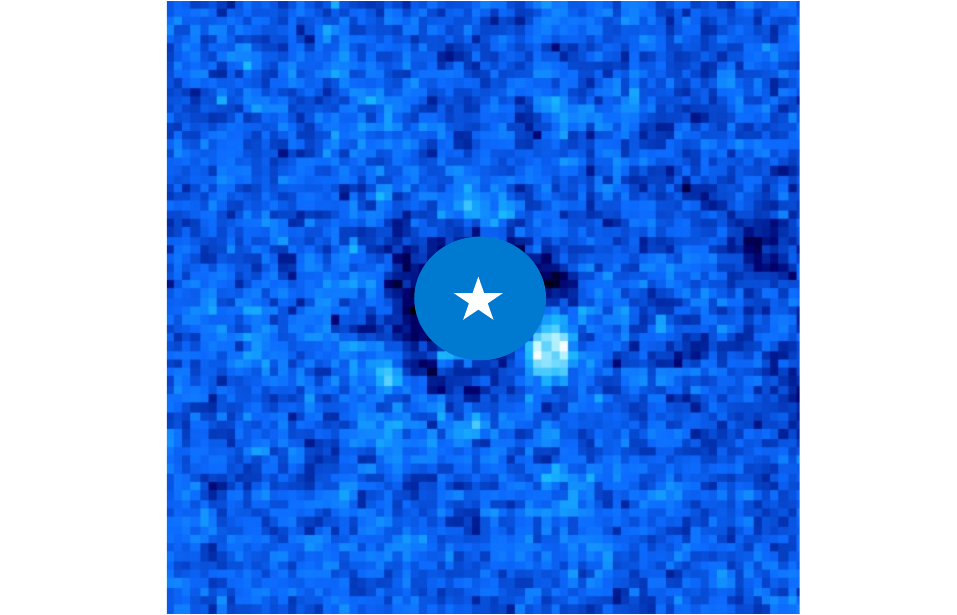}    \caption{ The feature to the lower right of the star’s location (the star has been subtracted) is due to persistence. Its hallmark “top-hat” appearance (a nearly flat, round flux profile) contrasts with the Gaussian shape expected for a true companion, indicating a false positive arising from a negative persistence imprint in the reference image.
}
    \label{FPexample}
\end{figure}

{An example from GO~6122 is shown in Figure~\ref{FPexample}. Here, we interpret the “source” to the lower right of the star’s position (the star itself, marked by a white star, has been subtracted) as persistence. Notably, the artifact exhibits a “top-hat” appearance, a nearly flat, round flux profile, characteristic of persistence, rather than the Gaussian profile expected for a true companion. Further analysis indeed confirmed that this feature is a negative artifact in the reference image, not a positive signal from a companion near the science target.}

To avoid false-positive exoplanet detections in MIRI high-contrast imaging, one should ensure that candidate planets do not appear at the separation corresponding to the stellar PSF offset between exposures if possible. In practice, RDI observations often involve multiple stellar PSFs and dithers, complicating (but not precluding) false-positive identification.

{ \subsection{Persistence Mitigation}}

{ Careful observation design can help avoid persistence issues. One straightforward mitigation strategy is simply to shorten the integration time on bright sources to avoid saturation, if science goals permit.}
While the saturation in this study was most severe at 7.7\,$\mu$m, saturation or near-saturation events can also occur at the longest MIRI wavelengths when observing bright stars (see \citealt{2025arXiv250515995B}). { However, saturation is often more severe at the shortest wavelengths when bright main-sequence stars are present. In such cases, it may be preferable to conduct MIRI imaging sequences in order of \textit{longest to shortest wavelength}} ideally with each using different dither positions (e.g., if band 1 uses cycling points $1-4$, then use points $5-8$ for band 2, etc).  { We implemented this dither approach for the final observation of Ross~154 in GO~6122 and found that it successfully mitigated persistence \citep[as well as brighter–fatter residuals that had previously arisen when using reference differential imaging instead of angular differential imaging on a slightly different brightness star;][]{2023A&A...680A..96A}}. 

{Conducting imaging sequences in order of longest to shortest wavelength is in tension with the current JWST documentation, which advises multi-band observations from short to long wavelengths because longer-wavelength filters have higher background flux that can induce persistence affecting subsequent shorter-wavelength exposures. Thus, our proposed reversal of the order should only be considered when integration times are relatively short and bright stars are present; otherwise, the default JWST documentation should be followed.
We note that later observations in program \#4403 began with the F2100W filter, following this strategy. Starting with the longest wavelength filter did not introduce any noticeable issues in the data nor did it appear to impact the absolute flux measurements of the white dwarfs in those systems.} 

When possible, space out in time the imaging of reference observations by a sufficient interval, allowing the persistence to decay ($\sim6+$ hours) while still maintaining a short enough gap to prevent significant wavefront error shifts in the observatory (linking the observations with a defined time gap rather than using a non-interruptible imaging sequence), {though this may not be practical within JWST's scheduling constraints, and using different dither positions may represent a more practical mitigation strategy.
}

\subsection{Impact of Persistence on Precision Photometric Measurements}

{MIRI's flux-deficit} persistence from bright targets can introduce significant errors in photometric measurements when it coincides with the location of a faint source observed at a later time. Faint sources emit less flux than the ``sky" background at the longest MIRI imaging wavelengths (e.g., 21\,$\mu$m; see the count rates in Figure \ref{fig:enter-label1}). Therefore, a 1-2\% {flux-deficit} in the background {due to persistence} can surpass the flux of the faint targets we aim to measure, such as an exoplanet or white dwarf. For example, infrared excess measurements of white dwarfs to detect companions or disks often require photometric precision at the level of a couple percent. However, persistence can contribute residual signals at the level of tens of percent of the white dwarf’s flux—far exceeding typical measurement uncertainties and potentially masking the presence of an unresolved planet or debris disk. 

When imaging faint sources (e.g., white dwarfs) with nearby bright companion stars, as in the data presented in this paper, care must be taken to ensure that the dither positions and their associated persistence trails do not overlap with the faint target. In this dataset, for instance, the white dwarfs targeted for infrared excess measurements unfortunately fell directly behind the persistence artifact of the bright star in the top two panels of Figure~\ref{fig:enter-label1} (dither 1, faint star near the upper right persistence spot), compromising the photometry.

\section{Summary}\label{summary}

In this work, we provided the first measurement of the persistence in MIRI imaging relative to the background at 21\,$\mu$m (F2100W). 
We measured the {flux-deficit due to} persistence to be $1.69 \pm 0.10$\% at 11.6\, minutes after the saturation event and $0.99 \pm 0.10$\% at 1.39\, hours, with the { the deficit} decaying by an order of magnitude after \(5.16^{+1.49}_{-0.94}\) hours. { While the exact decay timescale will vary with the severity of the saturating/near-saturating exposure and the pixel illumination history, our measurements suggest the effect becomes negligible on timescales of order $\sim$10\,h. Accordingly, for series of bright-star MIRI imaging where one wishes persistence to dissipate between visits, we recommend gaps of order $\sim$10\,h when feasible.}

As discussed in the previous section, {MIRI's flux-deficit persistence can contaminate photometric measurements and degrade PSF subtraction performance}, particularly small working angles. { Because the sky background is highest at longer MIRI wavelengths, and the flux-deficit persistence artifact manifests as a percent-level deficit relative to that background, the persistence appears more prominent in the longer-wavelength MIRI data (e.g., F1500W--F2550W, with F2550W most affected). One plausible contributor is that the higher zodiacal/thermal background, especially beyond $\sim$20\,$\mu$m, yields higher signal-to-noise on the background itself, making a percent-level deficit easier to discern; however, very deep short-wavelength data \citep[e.g., F770W, see ][]{2024A&A...689A...5D} with high-S/N backgrounds can also reveal the effect. We emphasize that detector persistence is not expected to be intrinsically wavelength dependent; rather, the apparent strength and morphology of the artifact in subsequent exposures can vary with wavelength via differences in background level (and, potentially, PSF and detector response).}
  Mitigation strategies to avoid persistence issues with MIRI high-contrast imaging are detailed in section \ref{discussion}.

As JWST continues to revolutionize our understanding of the mid-infrared universe, accurately characterizing detector effects like persistence is essential to maximizing its scientific potential. A deeper understanding of these artifacts will not only improve the reliability of current observations but also enhance the precision of future studies, ensuring that JWST achieves its full capability in detecting and analyzing faint astrophysical targets at the longest mid-infrared MIRI wavelengths.

 \section*{Acknowledgement}
{We thank the reviewer, Dan Dicken, for detailed and insightful comments that significantly improved this manuscript.} This work is based on observations made with the NASA/ESA/CSA James Webb Space Telescope. The data were obtained from the Mikulski Archive for Space Telescopes at the Space Telescope Science Institute, which is operated by the Association of Universities for Research in Astronomy, Inc., under NASA contract NAS 5-03127 for JWST. These observations are associated with program \#4403. This research has made use of the SIMBAD database,
operated at CDS, Strasbourg, France \citep{2000A&AS..143....9W}. 

\facilities JWST. {\it  The JWST data presented in this article were obtained from the Mikulski Archive for Space Telescopes (MAST) at the Space Telescope Science Institute. The specific observations analyzed can be accessed via \dataset[doi: 10.17909/cjkx-kp07]{https://doi.org/10.17909/cjkx-kp07}}

\software{ {\tt edmcmc.py} \citep{2021zndo...5599854V},  {\tt astro.py} \citep{2013A&A...558A..33A, 2018AJ....156..123A, 2022ApJ...935..167A}, {\tt numpy.py} \citep{5725236}, {\tt SAOImage DS9} \citep{2000ascl.soft03002S}}

\bibliography{main}{}
\bibliographystyle{aasjournal}
\end{document}